DO YOU KNOW WHERE TO LOOK
Harm H. Hollestelle, Amsterdam, The Netherlands

ABSTRACT

A group of line drawings in recent reports made by physics students during experimenting is investigated in order to describe the attitudes that can be related to the act of drawing. These attitudes represent time behavior and are related to how the experimenter basically perceives time. Also two concepts, one of desire, by Levinas, and the concept of constitution, by Husserl, are applied to describe two different notions of science. These notions are related to the question where the focus of the experiment is on: in the case of field work with responsibility for the infinite, and in the case of testing with responsibility for the un-distracted given. At least for this investigated group of drawings a meaningful correspondence can be shown to exist between the attitudes of the act of drawing and these notions of science.

1. INTRODUCTION

Drawings, images and analogies play a major role in the development of physics. Of the numerous studies that provide proof of this I mention M. Hesse's monograph [ref.1, and references therein] discussing the role of models and analogies in science. Other references can be found in [ref.2].
In this article recent (line-)drawings of an experimental situation are investigated. These drawings represent the experimental situation at the moment the experiment is performed. They include a part of the experimental set-up as far as it is regarded as essential and a part of the theory as far as it is considered relevant. The drawings will be regarded as models of both the experimental set-up as of the theoretical content of the experiment. That is, they are regarded as a model1 in the definition of M. Hesse [ref.3]. This means that besides positive and neutral analogies the drawing does not contain negative analogies of the experimental situation. A truly representative drawing will not contain elements that are contradictory to the experimental-setup or the theory. However, the artist is allowed to give an artistic interpretation of the elements chosen to be drawn. A drawn representation is not like a perfect copy, a drawing may magnify or neglect.
It is the experimenter (who at the same time is the artist) who decides at the moment of experimenting what part is essential and relevant. Besides this the models will also contain information from other sources from outside the experiment. By virtue of their construction these models embody an element of independence from both theory and set-up: it is because they are made of a mixture of elements including those of outside the original domain of investigation, that they maintain this partially independent status [ref.4]. They are drawings and because of this they obey rules of their own besides their representational function.

Models are instruments that can both structure and display measurement practices [ref.5]. They are often used for exploring or experimenting on a theory that is already in place. But also models are needed as instruments for exploring processes for which our theories do not give good accounts. In [ref.5] is given a survey of many possible situations in which models can be used as instruments. Drawings can have many of these functions.
Drawings also have a function in communicating the essence of an experiment. According to Polanyi it is craftwork that is the basis for the communication of the novel, private experience into the accepted public knowledge [ref.6, ref.7]. Craftwork in performing an experiment will show in the manner a drawing is made by the experimenter. This makes the drawing a source of possible communication. The intention of the experimenter and the motivation behind the experiment will be represented in the drawing. On the other hand the experimenter will be influenced during the act of drawing: this is not only a private activity, that can exist without experiencing the influence and demands of other physicists, who will be the future observers. Only think of the demand, drawing a truly readable and convincing sketch, makes on the artist.
Thus it is only natural to investigate drawings by physicists in order to establish a fundamental correspondence between the drawings on the one hand and the intention of the experiment, its meaning, on the other hand.

This article is only a study of a specific group of drawing, in relation to it only two different notions of science. The group of drawings consists of drawings in reports (or note-books) by physics students, made during experimenting. These are individual handmade line drawings. Some drawings are made with the help of a computer. Not meant are photographs of the experimental set-up or textbook drawings. The group of drawings is described in paragraph 2. Focused is on the act of drawing and the attitudes with which a drawing is made.
The meaning of an experiment is the intention with which it is performed. One of the main influences on this is the notion of science of the experimenter. Two notions are chosen as exemplary, without meaning with this that all other notions are dependent on these two. The two notions can be described by the terms field work and testing. Some experimenters will be performing the experiment with exactly one of these two notions as underlying motivation. Others will perform experiments with a mixture of, or a deviation from these two notions.
I will single out two drawings, from two reports, by different students, of the same experiment, to relate these drawings to the two above mentioned notions of science. A detailed description of the drawings will show how these notions of science work out. In order to do this the concepts of desire by Levinas and the concept of constitution by Husserl are applied.

## 2. ATTITUDES DURING THE ACT OF DRAWING

In this article use is made of the results of [ref.8]. Already in [ref.8] recent drawings in reports of physics student were studied. I will now summarize some results for further reference. Investigated were basic building block aspects of drawings that are related to the motor experiences of the experimenter. Those aspects were chosen that can be measured or counted, in order to be able to compare drawings. For instance chosen were aspects like some sizes of the drawings, that are related to the experience of reaching and distance. One major result was that large scale size of a drawing correlates inversely with the small scale size of the same drawing. This correlation was found to be true for the majority of drawings. Hence it was called a natural correlation. We will concentrate now mainly on this correlation.

As an explanation for this correlation it was argued that the larger a drawing is, the more the artist feels free to work on finer details. He or she feels free from the boundary, safe inside the larger space of the drawings' surface. This correlation is not strict. A small group of drawings was found that did not agree with this: these had relatively small values for both the large scale size and the small scale size.
The natural correlation can be visualized with the help of a concept/phenomenon called 'breath'. With this is meant that the flow of attention of the artist (or the observer, these flows are essentially the same, as is argued in [ref.8]) is compared with a series of sphere volumes. A larger drawing then relates to a flow of attention resembling a large number of small sphere volumes. A smaller drawing correlates with a flow of attention resembling a small number of large sphere volumes. When changing, as artist, from a large drawing to a smaller drawing, the total breath or total sphere volume remains the same. This constraint of constant total breath applies only to artists (and drawings) that follow the natural correlation. Other drawings with different correlation will have another concept of breath without this constraint. The concept of breath in terms of sphere volumes is just an aid for visualization. The word breath is chosen because, later in the text, this concept, as resembling the flow of attention, will be interpreted as the link between motor experience, and thinking and aiming, usually associated with breathing.

Drawings that follow the natural correlations can be divided in two clusters. One, cluster I, for the larger drawings with finer details, contains also other characteristics that correlate positively with the large scale size. Cluster II, for smaller drawings without finer details, contains characteristics that correlate inversely with large scale size.

The characteristics represent the attitudes of the artist with which a drawing is made. For instance cluster I contains the care and time attitude, defined as an attention flow with a large time span (large 'recurrence time') related to transversing the drawing as a whole. Cluster II contains the unity attitude, defined as an attention flow with a short time span (small 'local displacement time') related to transversing a standard detail of a drawing [ref.8]. Both of these attitudes are time behavior attitudes, i.e. they describe how the artists, when making the drawing, performs in time the strokes that make up the drawing. Other attitudes related for instance to distance and space (reaching), contact (touching) etc., can also be described.

It is possible to change from attitudes, however, when changing from one attitude to another the artist also changes the characteristics of the drawing and together with these the composition of breath changes likewise. As long as the artist stays within the natural correlations the constraint of constant total breath is effective.

In the following paragraphs it will be elaborated on that a change of attitude or breath is accompanied with a change of the notion of science of the artist. This will be done for the two attitudes mentioned above: the time and care attitude and the unity attitude will be treated in the next two paragraphs. Two real drawings are discussed in paragraph 5. A change of attitude will be described in the discussion section paragraph 6.

## 3. THE CARE AND TIME ATTITUDE AND FIELD WORK

Let us now in this paragraph focus on the attitude (called the care and time attitude) related to a large value for, global, recurrence time. In the next paragraph there will be discussed the attitude related to a small value for local displacement time (called the unity attitude).

A large value for the recurrence time means that it takes a relatively long time for the flow of attention (of the artist or equivalently the observer) to cover the drawing as a whole. This time value can be estimated for each line drawing with the help of a kinematic model that is explained in [ref.8].

Granting a large value to the recurrence time is interpreted as an attitude whereby the artist gives time and care to the drawing process. It is an attitude of patience. It is an attitude of waiting. There is no limit to the time devoted to the act of drawing in principle.

We will now try to connect with this attitude a notion of science that also has it's basis in patience and waiting. For this we refer to the concept of desire by E. Levinas [ref.9]. In [ref.9] M.-A. Lescourret states that according to Levinas "the philosophical quest for wisdom, for knowing, begins first with the wisdom of quest, of philia, of desire". To explain let me quote further: "We started with the burden of being and the question of the meaning of being. We end up with an understanding of the intelligibility of being according to a reorientation of sense, of the link between man and the world. It does not go from me to the world but in the reverse order from the world to me."

To understand this position one has to note that for Levinas (according to M.-A. Lescourret) "desire that links me to the other is not a matter of deliberate choice. Someone that is not me, I am linked to as it were 'in spite' of myself. The election is not my doing: Quite the opposite, I am elected to this election." And she goes on to explain Levinas: " Desire is free because it is not a matter of will and cannot be limited by my egotist gravitation. In desire I am free to myself and also free, not determined by the specific others, since my ignorant and innocent desire concerns alterity altogether beyond my will and wish."

Most clearly is the statement [ref.9]; "The object of desire – if ever desire needs an object – is clearly the undesirable, what cannot be desired because it can be neither named, nor called nor designated, but is what calls and names and designates the self in its passivity and responsibility."

This is a notion of science that is preparing for the unknown but passive because one does not know where to look. Formulated alternatively, it is responsible because one is passive and waiting. Only by waiting one may find what is unknown beforehand. One has to take care and time to be receptive and sensitive. One has to be open to the infinite.
This notion of science is a real and possible formulation of a practice of science as it is performed by many natural scientists. There are many experiments where the infinity of nature is explored to find new physical objects or manifestations. This is particularly so for field work in astronomy or biology. Field work is a common name designating work where a particular, unknown, field or area is covered and explored to find anything that has not been imagined beforehand. Being patience is a necessary condition to be successful during fieldwork. The care and time attitude is an attitude that comes close to describing how this is realized in general.

4. THE UNITY ATTITUDE AND TESTS

A small value for the local displacement time means that, as is explained in [ref.8], the time for transferring with ones attention a standard local distance in the drawing is short. One grasps, as it where, the drawing in a unity of short intervals, at once. This act is interpreted as the unity attitude, a time attitude where the artist draws the lines actively in a uniting thrust.
This is an attitude of activity and construction. A final and decisive result is meant to be arrived at in the experiment and there is hope for the emergence of truth in the drawing.

A notion of science that comes near to this unity attitude is one that constructs its data starting from as little as possible and using the most simple tools. For this notion we refer to the concept of constitution introduced by Husserl as the basis of phenomenology. In [ref.10] Husserl's concept of constitution is explained as follows. A subject, that is, in this case a physicist, can only know an object as phenomenon, as the object appears to be to the physicist, the subject. However the subject is not a passive receptor only, he or she constitutes, starting from the ego, the phenomenon as a product of the ego's content. This content is depending mostly on language, and time and space. Starting from the content of the ego, the phenomenon is actively given a meaning.
How are things in an scientific setting put together and experienced? Husserl [ref.11] in this context mentions the role of pure, perfect forms in the origin of geometry: "First to be singled out from the thing-shapes are surfaces – more or less 'smooth', more or less perfect surfaces; edges, more or less rough or fairly 'even'; in other words, more or less pure lines, angels, more or less perfect points; then, again, among the lines, for example, straight lines are especially preferred, and among the surfaces the even surfaces; for example, for practical purposes boards limited by even surfaces, straight lines, and points are preferred, whereas totally or partially curved surfaces are undesirable for many kinds of practical interests." And: "formations developed out of praxis and thought of in terms of gradual perfection, clearly serve only as a basis for a new sort of praxis out of which similarly named new constructions grow." "This new sort of construction will be a product arising out of an idealizing, spiritual act, .., and creates 'ideal objects'."

To rescue Husserl's phenomenology from solipsism in [ref.12] it is argued that for language to be a sufficient basis for constitution it is necessary that this language is not a private language but a real language. This language has to be communicated by other ego's, by others that already use language in a meaningful way. In this way the presence of others is secured [ref.12].

Images are, perhaps, not as general as language. However thinking in terms of images without words is possible. Everybody is familiar with the experience of 'being', when walking on the beach while the sun sets over the distant sea horizon. This is an experience that is just as strong as the Cartesian ego cogito, only without the words. Also images can be general in the way that an image can be non-specific like a portrait of some unknown person can stand for the face of the 'other' in general. Restricted to the area of drawings by physicists one may refer to standard textbook drawings like drawings of electrical circuits. Also images appear to nearly everybody. Nearly everybody is familiar with the image of the clear night sky with stars or the image of the grave. These are non-specific general images that occur to people just as frequently and generally as words do.
The above concept of constitution thus applies to both the use of words and the use of images when a phenomenon is given meaning. Likewise as with words, also with images meaning can be generated. Drawing thus means giving this drawing meaning by the act of drawing, by constituting this image.

This is a notion of science that is as active as possible, and that tries to be as less passive as possible. It is not preparing for the unknown to be received but it is actively pursuing meaning and aiming at a decisive result. This notion of science is responsible because it does not forget to look where it expects results. One knows where to look. And one expects results in the most direct and straightforward direction and movement. For this to be successful one has to have hope and decisiveness. Hope for the action of the experiment to be aiming correctly. Decisiveness for the action of the experiment to be aiming without being deflected. This notion of science is common among natural scientists when they perform 'yes or no' experiments, otherwise called tests. Tests are common experiments among physicists when they want to verify theoretical findings. These are experiments where the knowledge of nature is united and grasped at one time. Tests ask for a performance that is as short in time as possible, one should reach for the result as direct and as swift as possible. From these considerations one can conclude that the unity attitude is a very accurate description for the performance of tests and the related concept of constitution.

5. EXAMPLES FROM DRAWINGS BY PHYSICS STUDENTS

In [ref.8] a set of drawings made by students of physics for reports during experimenting was investigated. From this set of drawings, two are chosen as examples of how attitudes can be recognized from the drawings. The drawings are included at the end of this article as figure 1 and figure 2. Figure 1 represents a drawing that has a long global recurrence time, figure 2 represents a drawing that has a short local displacement time. Both drawings show the same experiment, that studies the interference of waves through different media, water and air. I will now give a description of the drawings and from the description I will choose the notion of science that corresponds with the treatment of the artist.

Figure 1 has much space, and much concern for details. There is an atmosphere of timeliness and peace, there is not much action in it. Notice the fine details of the footing of the apparatus for transmitting ('zender') and receiving ('detector') of the waves and the oscillator. This gives the objects a weight, a sensitivity to gravitation. The site were the actual measuring of the investigated phenomenon takes place is given the same or even less attention as all other sites of the experimental set-up. Nothing has escaped from the artists eye and everything has been depicted with responsibility for small details that otherwise would be neglected. There is responsibility for the infinity of the experimental situation. The experimenter does not choose were to look, when drawing. The experimenter does not show that he or she knows where to look because every part of the experimental situation could be giving decisive information. This is the strength of the drawing and the artist. There is interest in unpromising and perhaps boring or uninteresting data. The experiment may fail to give results along the expected line or measurements may ask for a time consuming treatment before giving results. This experimenter clearly is prepared for field work.

Figure 2 is very close and small of design and intense. There are no details at all, or rather the parts that make up the drawing are reduced to their essential bare form. There is a direct concentration, concentration on the site within the experimental set-up where the measurement (of the interference of waves) takes place. The experimental situation is reduced to what is necessary for imagining the actual measurement, the actual construction of the data. This experimenter knows where to look, and nothing is allowed to deflect from this. Notice the absence of everything outside the range of the two travelling waves. Only the experimental situation between the transmitter ('zender') and the receptor ('ontvanger') is drawn with emphasis on the part with the water column ('waterkolom') in the centre. The shape of the waves as they are drawn is corrected during drawing where necessary to secure that it truly resembles the physical reality. This time the interfering waves, that are at the heart of the experiment, are treated with all responsibility. The question the artist is asking is: will the waves show interference, will they or will they not behave as depicted in the drawing. This question asks for a direct and decisive answer. This experiment cannot fail to give a direct answer. The fact that the artist has been able to single out this question by means of a drawing attitude is the strength of the drawing and the artist. This experimenter has chosen an approach resembling that of a test experiment.

The figure 1 drawing has, because of the long global recurrence time, the care and time attitude. This corresponds as shown above with the notion of science as field work. The figure 2 drawing has, due to the short local displacement time, the unity attitude, which corresponds with the notion of science as testing. Both artists (both the experimenters) depict the same experiment, but their attitude is different and their notion, their meaning, is different. Both drawings have their own individual merits. Both drawings belong to reports of experimenters that were successful in the performance of this experiment. Also from an artist's view there is no reason to prefer one above the other in principle. All depends on the individual preferences of individual observers when one drawing is chosen as most appropriate for the experiment.

6. DISCUSSION

In the above two paragraphs two different attitudes possible for an artist making a line drawing are compared to two notions of science and of experimenting that are encountered in the natural sciences. The time and care attitude is compared with field work, the unity attitude with tests. These comparisons reveal close similarities between the attitudes and the notions. These similarities are not accidental: one cannot overlook that these attitudes of drawing are related and indeed similar from the start to the notions of science/experimenting.

The artist is the same person as the experimenter. Attitudes are not just a fashion of doing things. They are deeply rooted in the personality of the performer. As is explained in [ref.8] attitudes depend strongly on motor perception experiences. These are the fundaments for one's notions about space, time, weight, distance, presence etc. These notions again are on the basis of the questions the experimenter seeks to answer by the experiment, the meaning of the experiment. And they are on the basis of the attitudes of drawing when the experimenter wants to sketch the experimental set-up.

Thinking, in terms of words or in terms of images, originates in experiences. Words for instance can be created by the mind in reaction to an overwhelming experience [ref.13]. In [ref.13, pp. 28, 57, 90] Cassirer argues: "Before the intellectual work of conceiving and understanding of phenomena can set in, the work of naming must have preceded it. For it is this process which transforms the world of sense impression into a mental world, a world of ideas and meanings". "When the immediate intuition has been focused and, one might say, reduced to a single point, does the mythic or linguistic form emerge". "(Then) the mental view is not widened but compressed; it is distilled into a single point. Only by this process of distillation is the particular essence found and extracted which is to bear the special accent of 'significance'." Only a very concentrated attention on such an experience will be able to save the meaning of the experience in the memory and to do this a denotative word can be used and stored with it.

To illustrate his argument Cassirer gives the example of several African and Ural-Altaic languages [ref.13, pp. 158]. There, the words that are used for spatial relations are derived from words for concrete materials and especially from names of parts of the human body. The concept 'above' is denoted with the word for 'head' and the concept 'behind' is denoted with the word for 'back' etc. It is clear from these examples that words and motor experiences are related. Thinking, in terms of words as is usually the case, is related to motor experiences equally.
This view corresponds with the account given by Polanyi [ref.6] for in-dwelling as an act of focusing on the large scale outward experiences while the attention is retreated from the small scale bodily movements and sensitivity. Words for new experiences are then imagined and created while tacitly being assisted by the motor experiences. Also this view agrees with the explanation of the natural correlations in terms of larger drawings correlating positively with finer details. In-dwelling can then easily be generalized to apply to any flow of attention between the small scale and the large scale size of a drawing. It is then an activity that is a generator of meaning for the consciousness of the experimenter.

The example given by Cassirer does coincident very well with the account given by Kozel [ref. 14] of psychoanalytical and phenomenological reasoning concerning 'the real' (Lacan) or 'the flesh' (Merleau-Ponty) respectively, that both denote the pré-symbolic experience of the 'other'. This experience is without symbolism, imagination or language. The real is like the dark mist that surrounds us, our body, when we are not able to interpret experiences. It is like driving a car without windows. The flesh constitutes a social variant of the same thing, the same experience. The emphasis now is more on the contact properties and feeling. There is inside and outside of the body at the same time and Euclidian space seems to disintegrate.

A really new experience, that necessarily will be an overwhelming experience, will ask for a new vocabulary. This will ultimately have to be agreed upon within the community of the one sensitive to the experience, in this case that is the physical community. The community's sensitivity is anyhow an important basis for the experimenter's sensitivity to start with.

The relation between attitudes of drawing and the meaning of the experiment is the link element breath. Breath communicates between the theoretical motivation and the practical drawing, between the mind of the experimenter and the motor faculties of the artist. In this way breath creates meaning.
Both attitudes and notions of science originate from the same basis and are linked by the element breath. The word breath is not chosen accidently. In [ref.15] a survey of the usage of the word breath as it is related to the activity breathing during theatrical performance is given. Breathing is a physical/bodily motor activity performed to let air in and out of the lungs. Also it is a preliminary to speaking and language. In this last activity (logical) thinking finds its origin. Thus for a human being motor perception and thinking are naturally related and linked through breathing. In the same way the concept/phenomenon breath defined above relates drawing attitudes and the notions of science. Thus it is clear that changing an attitude of drawing will change a notion of science and vice versa.

Changing attitudes of drawing can be tried out more easily than changing a notion of science. Notions of science tend to be hidden behind the experiment, that is itself not easily changed without great costs. Drawing is visible immediately and asks only for pencil and paper. Not only the above sketched natural correlations between aspects of drawing are possible, also all kinds of deviations are open to exploration. The concept of breath in relation to drawing is very fluid. When the experimenter for instance wants to go beyond the constraint of constant total breath found for the natural correlations he or she could try drawing the experimental set-up in a unconventional way. Also changing constraints is easier, due to the arguments of Cassirer, when a deviating experience is encountered. This could be the case, when the artist chooses to draw something not related at all with the experiment.

Of course changing to a deviating correlation of aspects of drawing and a deviating concept of breath through drawing is only possible when it is recognized as a challenge. With this is meant one has to be, some way or other, open to the change. When during drawing a change manifests itself one has to recognize it and be willing to explore its possibilities.

Breath and breathing are based on and linked to a person's most basic experiences and notions. These are not easily 'overruled' by new experiences. Drawing could be helpful when a change in these concepts is needed. The description of both drawings and the act of drawing in terms of attitudes is shown to be meaningful and useful in this way. This is at least so for the drawings that belong to the area of reports and journals made by physicists during experimenting. Started was with the investigation of a group of line-drawings belonging to this area, made by students. These drawings are expected to stand model for drawings made by physicists in general.

Figure 1

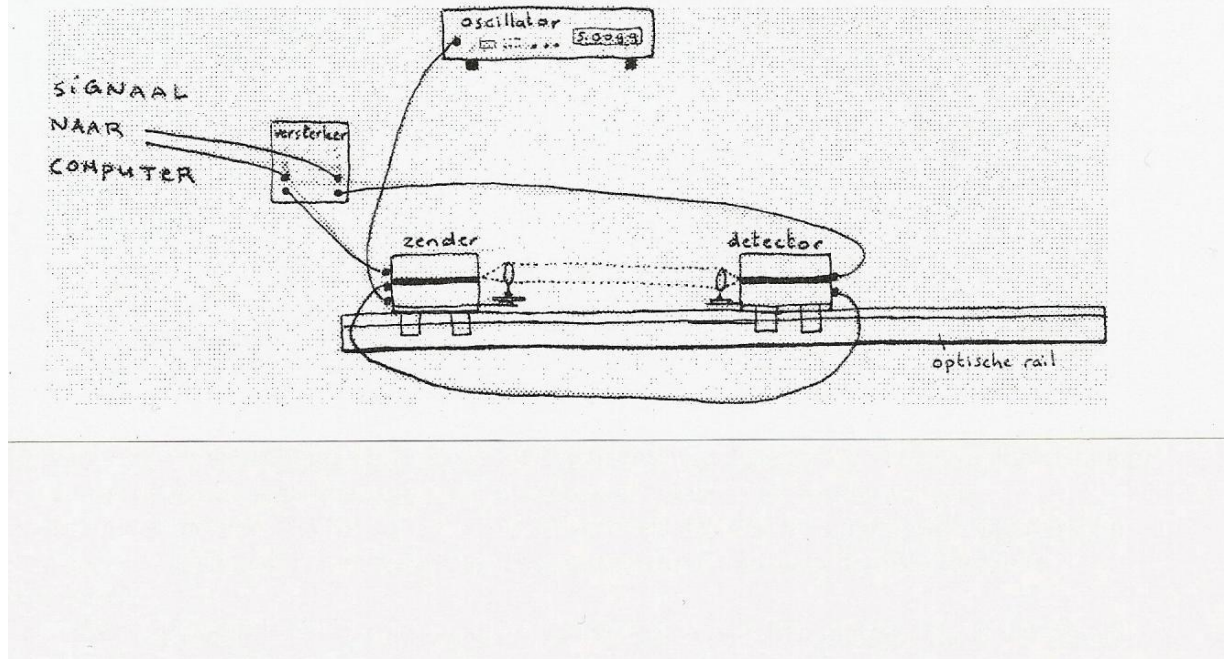

Een tekening van de opstelling:

Figure 2

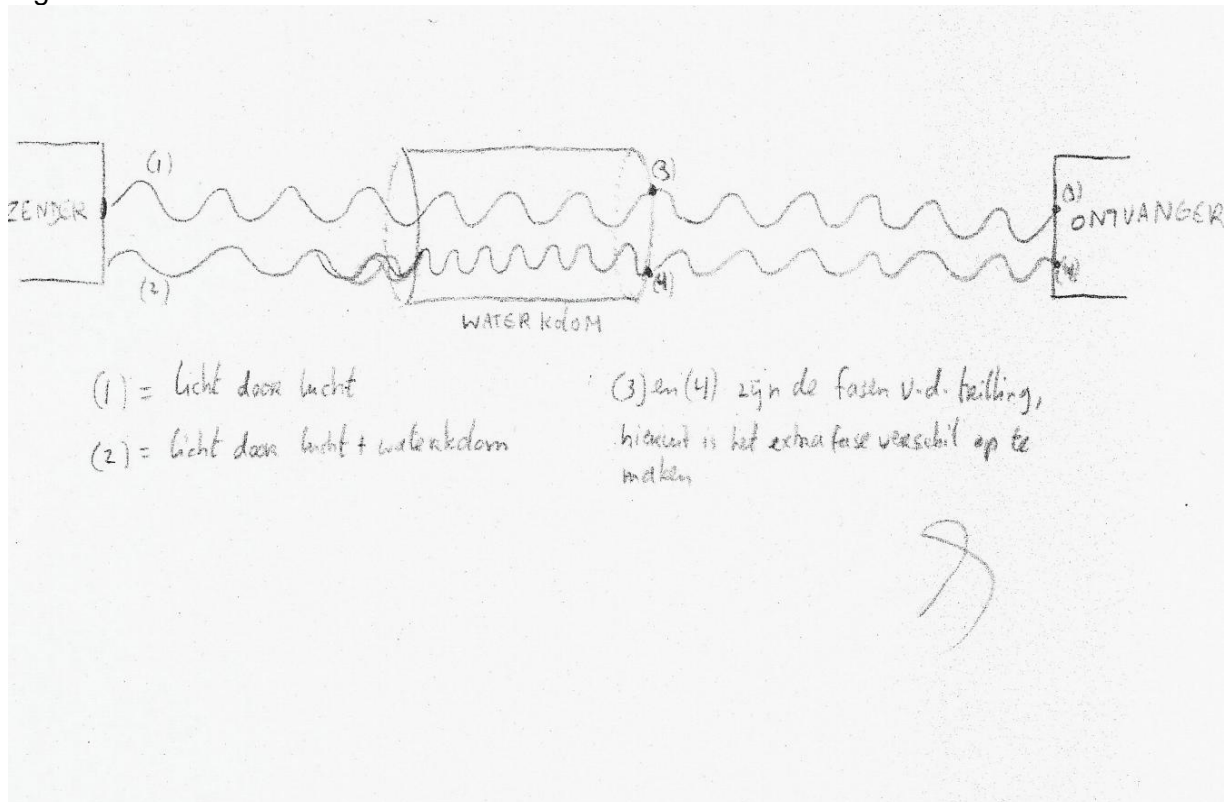

(1) = licht door lucht
(2) = licht door lucht + waterkolom

(3) en (4) zijn de fasen v.d. trilling, hieruit is het extra fase verschil op te maken